\documentclass[fleqn,10pt,]{wlscirep}
\usepackage[utf8]{inputenc}
\usepackage[T1]{fontenc}
\usepackage{amsmath}
\usepackage{braket}
\usepackage{algorithm}
\usepackage{algorithmic}

\newcommand{\argmin}{\mathop{\rm arg~min}\limits}

\title{Efficient partition of integer optimization problems with one-hot encoding}

\author[1,2,*]{Shuntaro Okada}
\author[2,3,4,5]{Masayuki Ohzeki}
\author[1]{Shinichiro Taguchi}
\affil[1]{Electronics R \& I Division, DENSO CORPORATION, Tokyo 103-6015, Japan}
\affil[2]{Graduate School of Information Sciences, Tohoku University, Sendai 980-8579, Japan}
\affil[3]{Institute of Innovative Research, Tokyo Institute of Technology, Yokohama 226-8503, Japan}
\affil[4]{Jij Inc., Tokyo 113-0033, Japan}
\affil[5]{Sigma-i Co. Ltd., Tokyo 108-0075, Japan}

\affil[*]{shuntaro\_okada@denso.co.jp}

\begin{abstract}
Quantum annealing is a heuristic algorithm for solving combinatorial optimization problems, and D-Wave Systems Inc. has developed hardware for implementing this algorithm.
The current version of the D-Wave quantum annealer can solve unconstrained binary optimization problems with a limited number of binary variables, although cost functions of many practical problems are defined by a large number of integer variables.
To solve these problems with the quantum annealer, the integer variables are generally binarized with one-hot encoding, and the binarized problem is partitioned into small subproblems.
However, the entire search space of the binarized problem is considerably extended compared to that of the original integer problem and is dominated by unfeasible solutions.
Therefore, to efficiently solve large optimization problems with one-hot encoding, partitioning methods that extract subproblems with as many feasible solutions as possible are required.
We propose two partitioning methods and demonstrate that better solutions are obtained using the methods proposed in this study.
\end{abstract}

\begin{document}

\flushbottom
\maketitle
%
%
\thispagestyle{empty}
\section*{Introduction}
Combinatorial optimization problems, i.e., the minimization of cost functions with discrete variables, have significant real-world applications.
Generally, the cost function of a combinatorial optimization problem can be mapped to the Hamiltonian of a classical Ising model \cite{Ising_mapping}.
Simulated annealing (SA) \cite{SA_original} is a classical heuristic algorithm that searches the ground states of a Hamiltonian, exploiting thermal fluctuations to escape local minima.
In contrast to SA, quantum annealing (QA) \cite{QA_original}, which is strongly related to the adiabatic quantum computation \cite{AQC_original}, escapes the local minima through the tunneling effects induced by quantum fluctuations.
Whether the quantum effects accelerate the computation of searching ground states is one of the main topics of research, and numerous studies have been conducted on this topic \cite{QA_SA_compare1, QA_SA_compare2, QA_SA_compare3, QA_SA_compare4, QA_SA_compare5, QA_SA_compare6}.
Recently, D-Wave Systems Inc. developed a commercial QA machine based on superconducting flux qubits \cite{D-Wave_machine}.
Experimental studies using QA machines have been performed to compare the performance of QA with that of SA \cite{D-Wave_compare1, D-Wave_compare2, D-Wave_compare3}  and to demonstrate the applicability of QA machines to practical problems
\cite{D-Wave_application1, D-Wave_application2, D-Wave_application3, D-Wave_application4, D-Wave_application5, D-Wave_application6, D-Wave_application7, D-Wave_application8, D-Wave_application9, D-Wave_application10, D-Wave_application11, D-Wave_application12, D-Wave_application13, D-Wave_application14, D-Wave_application15, D-Wave_application16, D-Wave_application17}.

The generic form of a time-dependent Hamiltonian in QA is as follows:
\begin{equation}
\hat{H} \left( t \right) = A \left( t \right) \hat{H}_{\mathrm{q}} + B \left( t \right) \hat{H}_0, \label{eq:QA_Hamiltonian}
\end{equation}
where $\hat{H}_0$ is the classical Hamiltonian which represents the cost function to be minimized,
and $\hat{H}_{\rm{q}}$ is the quantum fluctuation term for which the ground state is trivial.
At the beginning of QA, the coefficients of the time-dependent Hamiltonian are set to $A(0)=1$ and $B(0)=0$, and the system is in the trivial ground state determined by $\hat{H}_{\mathrm{q}}$.
At the end of QA, the coefficients are set to $A(\tau)=0$ and $B(\tau)=1$, where $\tau$ is the annealing time.
The system evolves according to the Schr\"{o}dinger equation:
\begin{equation}
i \frac{d}{dt} \ket{\psi(t)} = \hat{H} \left( t \right) \ket{\psi(t)},
\end{equation}
where $\ket{\psi(t)}$ is a state vector of the system and $\hbar$ is set to $1$ for simplicity.
According to the adiabatic theorem \cite{adiabatic_condition}, the system will remain close to the instantaneous ground state of the time-dependent Hamiltonian if it changes sufficiently slowly.
Thus, by setting the annealing time $\tau$ large enough, we can obtain the ground state of the classical Hamiltonian $\hat{H}_0$, which represents the optimal solution.

The current version of the D-Wave quantum annealer (D-Wave 2000Q) implements QA with a transverse magnetic field:
\begin{equation}
\hat{H}_{\mathrm{q}} = - \sum_{i=1}^{N_{\mathrm{q}}} \hat{\sigma}_{i}^{(x)},
\end{equation}
where $N_{\mathrm{q}}$ represents the total number of qubits.
A cost function that can be handled by the D-Wave quantum annealer is as follows:
\begin{equation}
\hat{H}_{0} = \sum_{(i,j) \in \mathrm{chimera}} J_{ij} \hat{\sigma}_{i}^{(z)} \hat{\sigma}_{j}^{(z)} + \sum_{i=1}^{N_{\mathrm{q}}} h_{i} \hat{\sigma}_{i}^{(z)},
\end{equation}
where the interactions between qubits are restricted to the Chimera graph, that is constructed as an $M \times N$ grid of complete bipartite graphs $K_{L,L}$ \cite{chimera_architecture}.
Although the Chimera graph for D-Wave 2000Q is $(M, N, L) = (16, 16, 4)$, the number of operable qubits is less than $N_{\mathrm{q}} = 2MNL = 2048$ because of defects in the qubits and connectivities.

Due to the limitation of the number of available qubits, we cannot solve large optimization problems directly using the D-Wave quantum annealer.
In real settings, large problems are partitioned into subproblems that can be handled by the quantum annealer.
The subproblems are iteratively optimized by the quantum annealer, and the optimization result is used to improve the current solution \cite{qbsolv, hybrid1, hybrid2}.
A cluster of spins in the subproblem are simultaneously updated in this scheme,
and this iterative method is considered as one of the large-neighborhood local search algorithms \cite{large-neighborhood}.
Although such algorithms can be performed using classical computers, subproblems are basically restricted to tree structures that are solvable in polynomial time by belief propagation or dynamic programming \cite{tree_partition1, tree_partition2, tree_partition3, tree_partition4}.
Therefore, employing the quantum annealer is considered to be advantageous if it can solve subproblems with many closed loops more efficiently than classical algorithms. 
Furthermore, it is conjectured that, for improving solution accuracy, solving as large subproblems as possible is essential.
The size of subproblems that can be embedded into the quantum annealer strongly depends on the quality of the minor embedding, particularly for problems that have a small number of interactions.
Because subproblems must be iteratively embedded, fast algorithms to embed larger subproblems are required for exploiting the potential of the quantum annealer.
While it is reasonable to employ a complete-graph embedding \cite{Comp_Embed1, Comp_Embed2, Comp_Embed3} for problems with dense interactions,
the subproblem-embedding algorithm, which we developed in a previous study \cite{subproblem_embed}, might be effective for improving solution accuracy of sparse problems.

\begin{figure}
	\centering
	\includegraphics[width = 1.0\columnwidth]{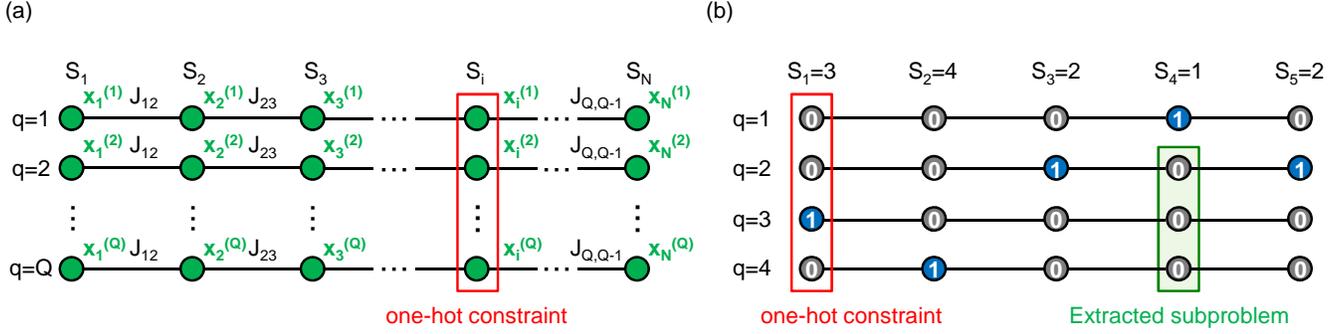}
	\caption{(a) Problem graph of Eq. (\ref{eq:1D_potts_OH}). Vertices and edges represent binary variables $x_{i}^{(q)}$ and interactions between them, respectively.
	Although the penalty term generates fully connected vertical interactions between $x_{i}^{(q)}$ and $x_{i}^{(q')}$, these are not shown for simplicity. $Q$ binary variables $\{ x^{(q)}_{i} \}_{q=1, 2, ..., Q}$ are assigned to each $S_{i}$, and the total number of binary variables is $NQ$.
	Although there exist $2^{NQ}$ configurations of the binary variables, only $Q^{N}$ configurations satisfy the one-hot constraint.
	(b) An example of an undesirable partition. Binary variables filled blue represent the tentatively selected component in the current solution.
	Better feasible solutions cannot be explored by optimizing the extracted subproblem whose binary variables are enclosed by the green rectangle.}
	\label{fig:one-hot_graph}
\end{figure}

Moreover, the quantum annealer requires that the cost function is represented in the form of a quadratic unconstrained binary optimization problem (QUBO) or Ising model, although many cost functions in practical problems are defined by integer variables.
Generally, the binarization of the integer variables is achieved using one-hot encoding \cite{Ising_mapping}.
For example, the following integer optimization problem with $N$ integer variables $\{ S_{i} \}_{i=1, 2, ..., N}$:
\begin{equation}
\argmin_{\{ S_{i} \}} \sum_{i=1}^{N-1} J_{i,i+1} \delta \left( S_{i}, S_{i+1} \right),  \label{eq:1D_potts}
\end{equation}
where $S_{i} \in ( 1, 2, ..., Q )$, $Q$ is the number of components, $J_{i,i+1}$ is an interaction between $S_{i}$ and $S_{i+1}$, and $\delta$ denotes the Kronecker delta function, is rewritten as
\begin{equation}
\argmin_{\{ x^{(q)}_{i} \}} \sum_{i=1}^{N-1} J_{i,i+1} \sum_{q=1}^{Q} x^{(q)}_{i} x^{(q)}_{i+1} \ \  \mathrm{s.t.} \ \  \sum_{q=1}^{Q} x^{(q)}_{i} = 1,
\end{equation}
by one-hot encoding.
Here, $x^{(q)}_{i} \in ( 0, 1 )$ is a binary variable that is assigned to the component $q$ of $S_{i}$, $x^{(q)}_{i}=1$ indicates that the component $q$ is selected for $S_{i}$, and feasible solutions are constrained to configurations in which exactly one component is selected for each $S_{i}$.
Subsequently, a penalty term is introduced to obtain the following unconstrained form:
\begin{equation}
H_{0} = \sum_{i=1}^{N-1} J_{i,i+1} \sum_{q=1}^{Q} x^{(q)}_{i} x^{(q)}_{i+1} + \lambda \sum_{i=1}^{N} \left( \sum_{q=1}^{Q} x^{(q)}_{i} - 1 \right)^{2},  \label{eq:1D_potts_OH}
\end{equation}
where the second term depicts the penalty term introduced to extract feasible solutions that satisfy the constraint $\sum_{q=1}^{Q} x^{(q)}_{i} = 1$, which we call "one-hot constraint", and the parameter $\lambda$ controls the strength of the penalty term.
By setting the parameter $\lambda$ to a sufficiently large value, ground states of the original integer optimization problem [Eq. (\ref{eq:1D_potts})] are correctly encoded.
However, the performance of the D-Wave quantum annealer is significantly affected by noise and intrinsic control errors when a needlessly large $\lambda$ is used.
Hence, to obtain high accurate solutions, we must explore an appropriate value of $\lambda$, which is one of the most tedious tasks for the optimization under the one-hot constraint.
Moreover, the intire search space of the binarized optimization problem [Eq. (\ref{eq:1D_potts_OH})] is dominated by unfeasible solutions.
Figure \ref{fig:one-hot_graph}(a) shows the problem graph of Eq. (\ref{eq:1D_potts_OH}), whose vertices and edges represent binary variables and interactions between them, respectively.
$Q$ binary variables $\{ x^{(q)}_{i} \}_{q=1, 2, ..., Q}$ are assigned to each $S_{i}$, and the total number of binary variables is $NQ$.
While the number of configurations of binary variables is $2^{NQ}$, the number of feasible solutions is only $Q^{N}$.
Therefore, to efficiently solve large optimization problems under the one-hot constraint using the quantum annealer, partitioning methods are required to extract subproblems with as many feasible solutions as possible.
A simple example of an undesirable partition is depicted in Fig. \ref{fig:one-hot_graph}(b).
Assume that one hopes to improve the current solution shown in Fig. \ref{fig:one-hot_graph}(b) and that the three binary variables enclosed by the green rectangle are extracted as the subproblem.
In this case, better feasible solutions cannot be explored by optimizing the subproblem as only the current solution in the subproblem satisfies the one-hot constraint.
To the best of our knowledge, the partitioning method proposed in the literature \cite{D-Wave_application17} is the first one focusing on the one-hot constraint.
This method is applicable to the double-constrained problems, $\sum_{q} x^{(q)}_{i} = 1$ and $\sum_{i} x^{(q)}_{i} = 1$ such as the assignment problem and the traveling salesman problem.
However, extracted subproblems still contain unfeasible solutions, for which the parameter $\lambda$ has to be adjusted.
In this study, we propose two partitioning methods applicable to the problems whose cost function involves a single one-hot constraint, as shown in Eq. (\ref{eq:1D_potts_OH}).
The first method is similar to the previously developed method \cite{D-Wave_application17}, while the other method extracts subproblems comprising only feasible solutions and does not require adjusting the parameter $\lambda$.
The performance of the proposed methods is assessed for several Potts models, which are generalized Ising models whose cost function is defined by integer variables \cite{The_Potts_model}.
We demonstrate that better solutions are efficiently obtained using the proposed methods.


\section*{Results}
In this section, we propose efficient partitioning methods for solving large optimization problems under the one-hot constraint, and assess the performance of proposed methods for several Potts models.

\subsection*{Proposed methods}
\begin{figure}
	\centering
	\includegraphics[width = 1.0\columnwidth]{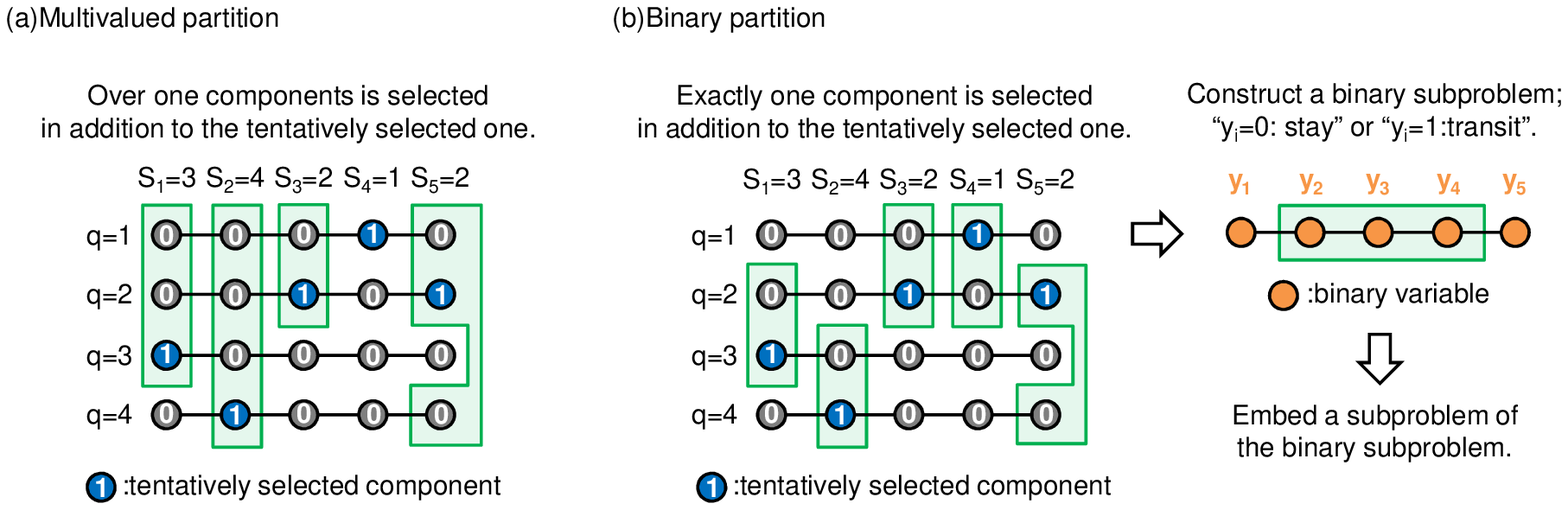}
	\caption{Proposed methods for efficient solutions of large optimization problems under the one-hot constraint.}
	\label{fig:proposed_methods}
\end{figure}

We propose two partitioning methods: one is the multivalued partition, and the other is the binary partition.
These methods are summarized in Fig. \ref{fig:proposed_methods}.
Both methods extract a subproblem that involves binary variables assigned to the tentatively selected components for each $S_{i}$.
The resulting subproblems include feasible solutions other than the current feasible solution.

The multivalued partition extracts a subproblem with two or more components for each $S_{i}$, as shown in Fig. \ref{fig:proposed_methods}(a).
In addition to the tentatively selected component, one or more components are randomly selected for each $S_{i}$, and a subproblem that comprises the binary variables assigned to the selected components is extracted.
The extracted subproblem involves feasible solutions other than the current solution, and the randomly selected components are explored for each $S_{i}$ by optimizing the subproblem.
However, the extracted subproblem still contains unfeasible solutions, and the penalty term remains in the cost function of the subproblem.
This partitioning method is similar to that developed in the literature \cite{D-Wave_application17}.
While the extracted subproblems are embedded usnig complete-graph embedding in the literature \cite{D-Wave_application17}, we employed the subproblem-embedding algorithm, which we developed in a previous study \cite{subproblem_embed}.
Details on how to achieve a multivalued partition using the subproblem-embedding algorithm are explained in the Methods section.

The binary partition is summarized in Fig. \ref{fig:proposed_methods}(b).
In addition to the tentatively selected component, the binary partition randomly selects exactly one component for each $S_{i}$.
Subsequently, new binary variables $\{ y_{i} \}_{i=1, 2, ..., N}$ that represent "stay in the tentatively selected component ($y_{i}=0$)" or "transit to the randomly selected component ($y_{i}=1$)" are introduced for each $S_{i}$, and a binary subproblem is constructed whose cost function is defined by $\{ y_{i} \}_{i=1, 2, ..., N}$.
The cost function of the binary subproblem is derived in the Methods section.
After that, a subproblem of the binary subproblem is embedded into the D-Wave quantum annealer.
Here, the cost function of the binary subproblem does not involve the penalty term because all solutions in the binary subproblem are feasible.
Therefore, the binary partition does not require adjusting the parameter $\lambda$.
Moreover, a larger number of binary variables can be embedded into the D-Wave quantum annealer because the penalty term, which generates fully connected interactions between $x^{(q)}_{i}$ and $x^{(q')}_{i}$, is not involved.
Consequently, the number of feasible solutions involved in the embedded subproblem is considerably increased using the binary partition.
The binary subproblem can be regarded as one of the simplest cases of the optimization under the half-hot constraint \cite{half_hot}.
The half-hot constraint is proposed to avoid the difficulty caused by the longitudinal magnetic field of the penalty term.
This difficulty is avoidable using the binary partition, which might contribute to improving solution accuracy.
A disadvantage of the binary partition is that only two components are considered for each integer variable.
As shown in the next subsection, this leads to poor performance for the ferromagnetic Potts model.

\subsection*{Assessing performance}
\begin{figure}
	\centering
	\includegraphics[width = 1.0\columnwidth]{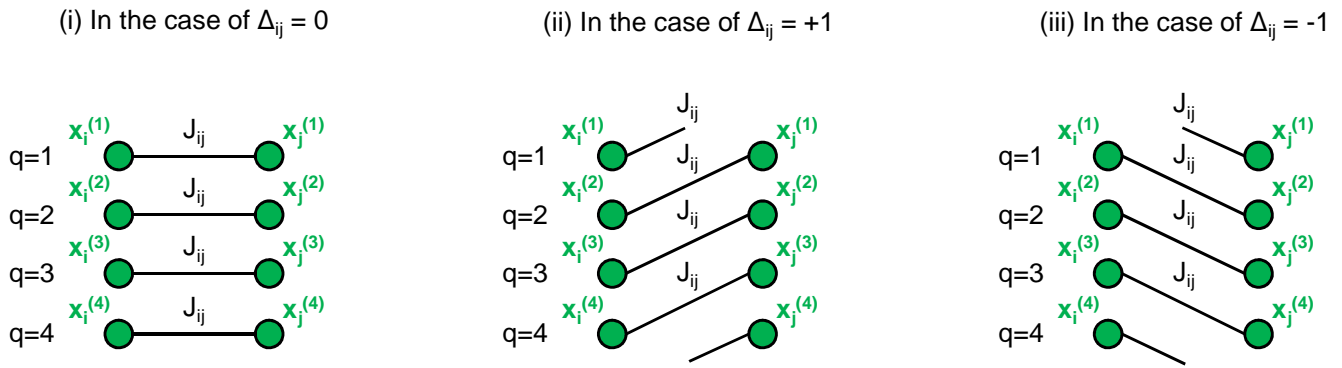}
	\caption{Local interactions generated by the first term of Eq. (\ref{eq:assess_one-hot}). $\Delta_{ij} \neq 0$ causes the interactions between different components.}
	\label{fig:local_interaction}
\end{figure}

The performance of the proposed methods is evaluated for four types of Potts models on the cubic lattice with $10 \times 10 \times 10$ integer variables, namely, the ferromagnetic, anti-ferromagnetic, Potts glass \cite{Potts_glass} and Potts gause glass  \cite{Potts_gauge_glass, chiral_Potts} models.
The cost function is given by
\begin{equation}
H_{0} = \sum_{<i,j>} J_{ij} \delta \left( S_{i}, S_{j} + \Delta_{ij} \right),
\end{equation}
where $Q$ is set to $4$, $S_{i} \in ( 1, 2, 3, 4 )$, $\Delta_{ij} \in ( 0, \pm 1 )$, $J_{ij}$ represents the interaction between the nearest neighbors on the cubic lattice, and $\delta$ denotes the Kronecker delta function.
The cost function is represented as the QUBO form using the one-hot constraint as follows:
\begin{equation}
H_{0} = \sum_{<i,j>} J_{ij} \sum_{q=1}^{4} x^{(q)}_{i} x^{(q-\Delta_{ij})}_{j} + \lambda \sum_{i=1}^{1000} \left( \sum_{q=1}^{4} x^{(q)}_{i} - 1 \right)^{2}.  \label{eq:assess_one-hot}
\end{equation}
The parameters $J_{ij}$ and $\Delta_{ij}$ in each model are shown in Table \ref{tab:parameter_setting}.
\begin{table}
\begin{center}
\caption{Parameter settings}
\begin{tabular}{ccc}  \hline
Model & $J_{ij}$ & $\Delta_{ij}$  \\  \hline
Ferromagnetic Potts model & $-1$ & $0$  \\  
Anti-ferromagnetic Potts model & $+1$ & $0$  \\  
Potts glass model & $+1(50\%)$ or $-1(50\%)$ & 0  \\  
Potts gauge glass model & $-1$ & $0(50\%)$ or $+1(25\%)$ or $-1(25\%)$  \\  \hline
\end{tabular}
\label{tab:parameter_setting}
\end{center}
\end{table}
$\Delta_{ij} \neq 0$ generates interactions between different components in the Potts gauge glass model.
The local interactions generated by the first term of Eq. (\ref{eq:assess_one-hot}) are shown in Fig. \ref{fig:local_interaction}.
\begin{figure}
	\centering
	\includegraphics[width = 1.0\columnwidth]{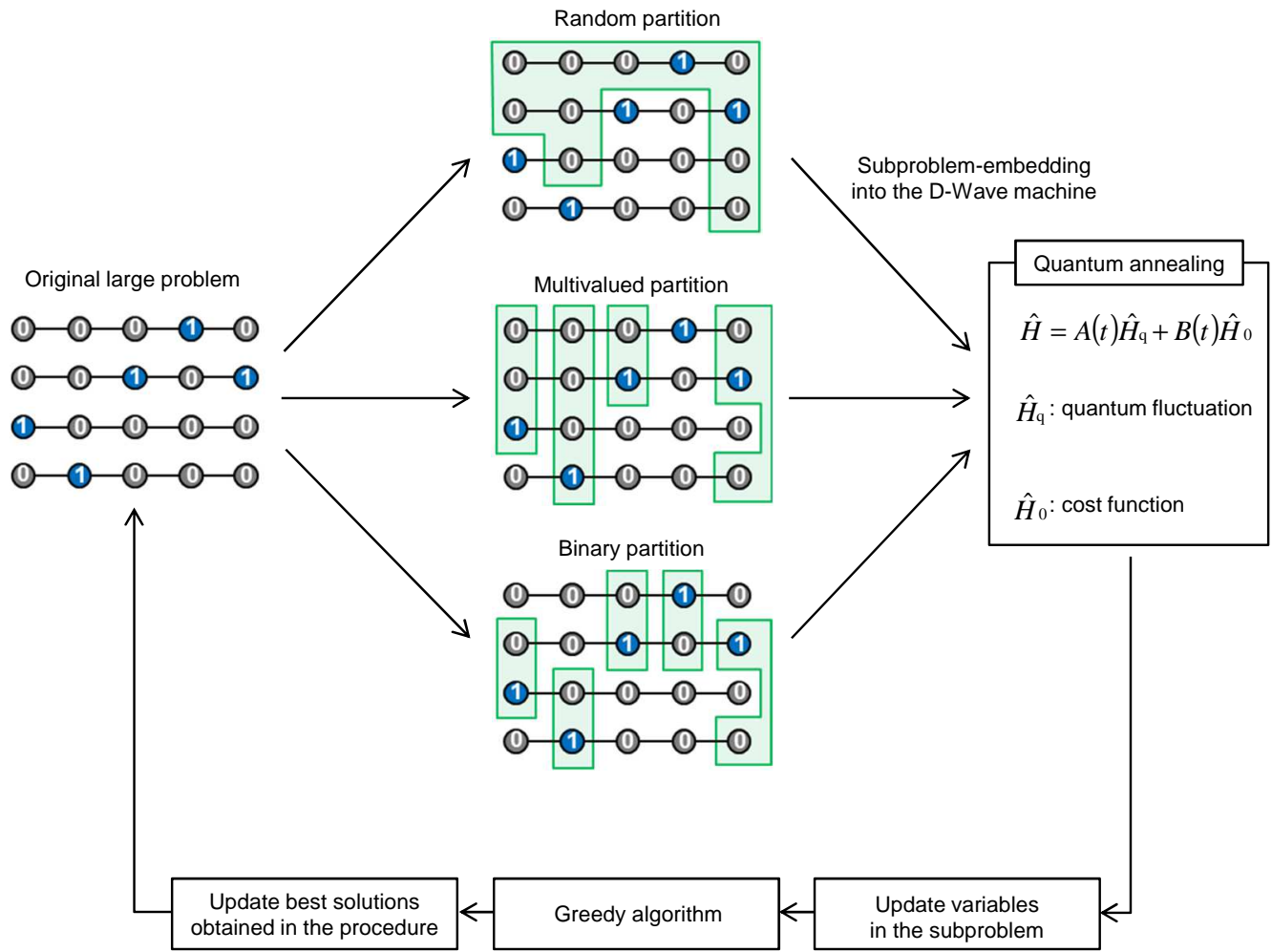}
	\caption{Optimization process demonstrated in this study. The solution accuracy is evaluated for the three partitioning methods.}
	\label{fig:opt_process}
\end{figure}
While the ground states of the ferromagnetic and anti-ferromagnetic Potts models are trivial,
it is generally difficult to obtain those of the Potts glass and Potts gauge glass models because of frustrations.

The optimization process demonstrated in this study is shown in Fig. \ref{fig:opt_process}.
The original large problem is partitioned using three partitioning methods: the random, multivalued and binary partitions.
The random partition does not adress whether an extracted subproblem contains feasible solutions for each $S_{i}$ or not.
The subproblem-embedding algorithm proposed in the literature \cite{subproblem_embed} is used for embedding a subproblem into the D-Wave quantum annealer.
After optimizing the embedded subproblem using the D-Wave quantum annealer, the variables in the subproblem are replaced to the best solution among the $1,000$ solutions obtained using the quantum annealer.
Subsequently, a greedy algorithm is executed by a conventional digital computer to get to exact (local) minima.
In this greedy algorithm, an integer variable $S_{i}$ is randomly selected, and the tentatively selected component is replaced to that which minimizes the local energy with respect to the selected integer variable $S_{i}$.
We complete refining the current solution when all local energies are minimized.
Finally, the best solution obtained in the procedure is updated.
These processes are iterated, and we compare the solution accuracy for the three partitioning methods.

\begin{figure}
	\centering
	\includegraphics[width = 1.0\columnwidth]{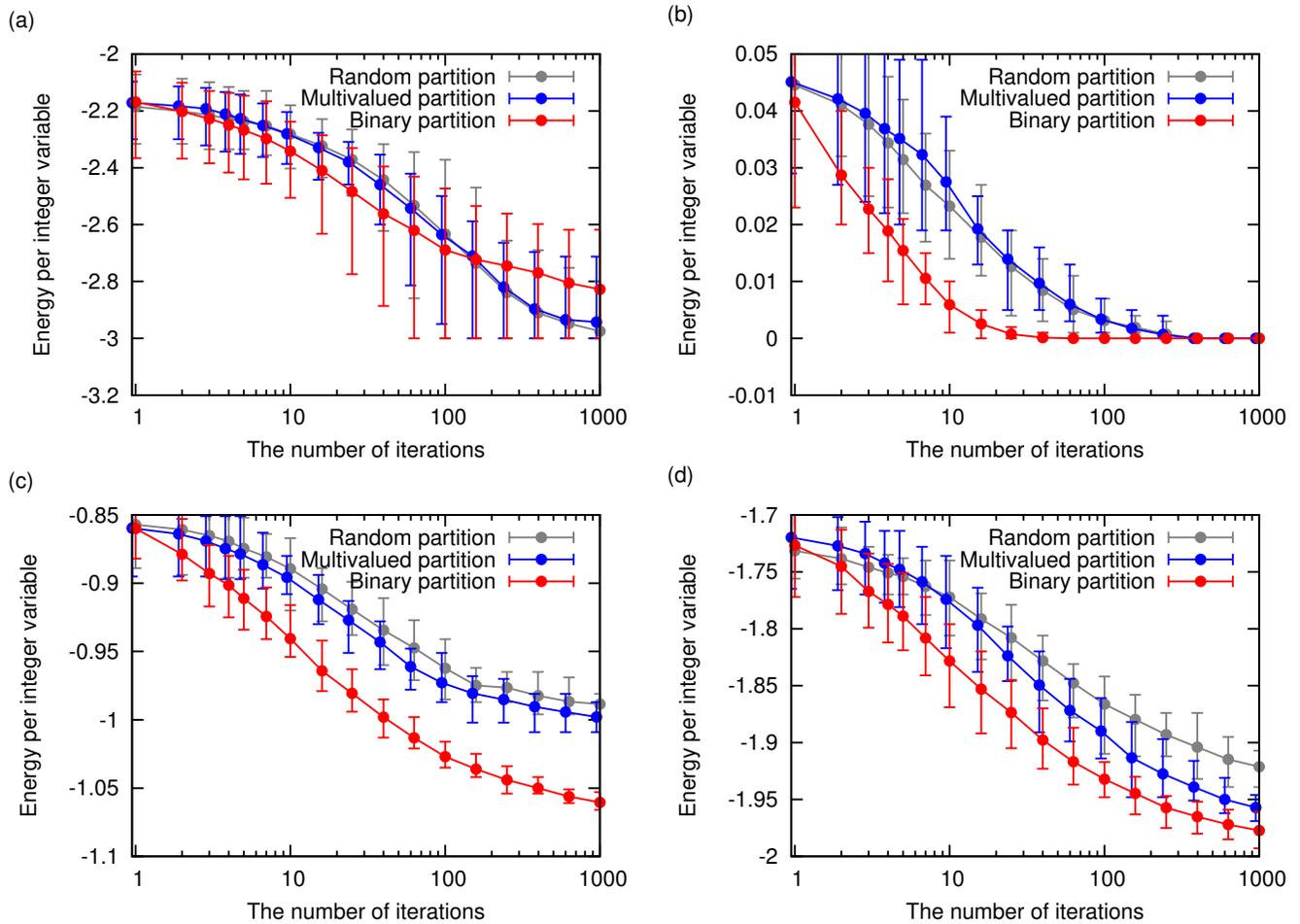}
	\caption{Energies obtained using the three partitioning methods. The average, maximum, and minimum energies for 16 trials are plotted. (a) Ferromagnetic Potts model. (b) Anti-ferromagnetic Potts model. (c) Potts glass model. (d) Potts gauge glass model.}
	\label{fig:result_energy}
\end{figure}

The obtained energies by the three partitioning methods are shown in Fig. \ref{fig:result_energy}.
The average, maximum, and minimum energies for 16 trials are plotted, and the same $16$ initial states are used for each partitioning method.
The horizontal axis represents the number of iterations, which is the number of subproblem optimizations performed by the D-Wave quantum annealer.
The plot for the multivalued partition is slightly shifted to the left to avoid the overlap between other plots.
Figures \ref{fig:result_energy}(a) and (b) show the obtained energies for the ferromagnetic and anti-ferromagnetic Potts models, respectively.
The ground states of these models are trivial, and the minimum energy is $-3$ and $0$ for the ferromagnetic and anti-ferromagnetic Potts models, respectively.
Although the multivalued partition is expected to solve large optimization problems more efficiently than the random partition, the performance of the random and multivalued partitions are almost the same.
The performance of the binary partition is different from the other methods; while it is the worst for the ferromagnetic Potts model, it is best for the anti-ferromagnetic Potts model.
Figures \ref{fig:result_energy}(c) and (d) show the obtained energies for the Potts glass and Potts gauge glass models, respectively.
As expected, better solutions are obtained with a smaller number of iterations using multivalued partition compared to random partition, particularly for the Potts gauge glass model.
The binary partition shows the best performance among the three partitioning methods for both the Potts glass and Potts gauge glass models.

\section*{Discussion}
In this section, we discuss differences among the three partitioning methods.
The following three questions arise from the results in the previous section.
\begin{itemize}
\item Why is the multivalued partition not superior to the random partition for the ferromagnetic and anti-ferromagnetic Potts models?

\item Why is the performance of the binary partition the worst for the ferromagnetic Potts model?

\item Why does the binary partition exhibit the best performance except for the ferromagnetic Potts model?
\end{itemize}

\begin{figure}
	\centering
	\includegraphics[width = 1.0\columnwidth]{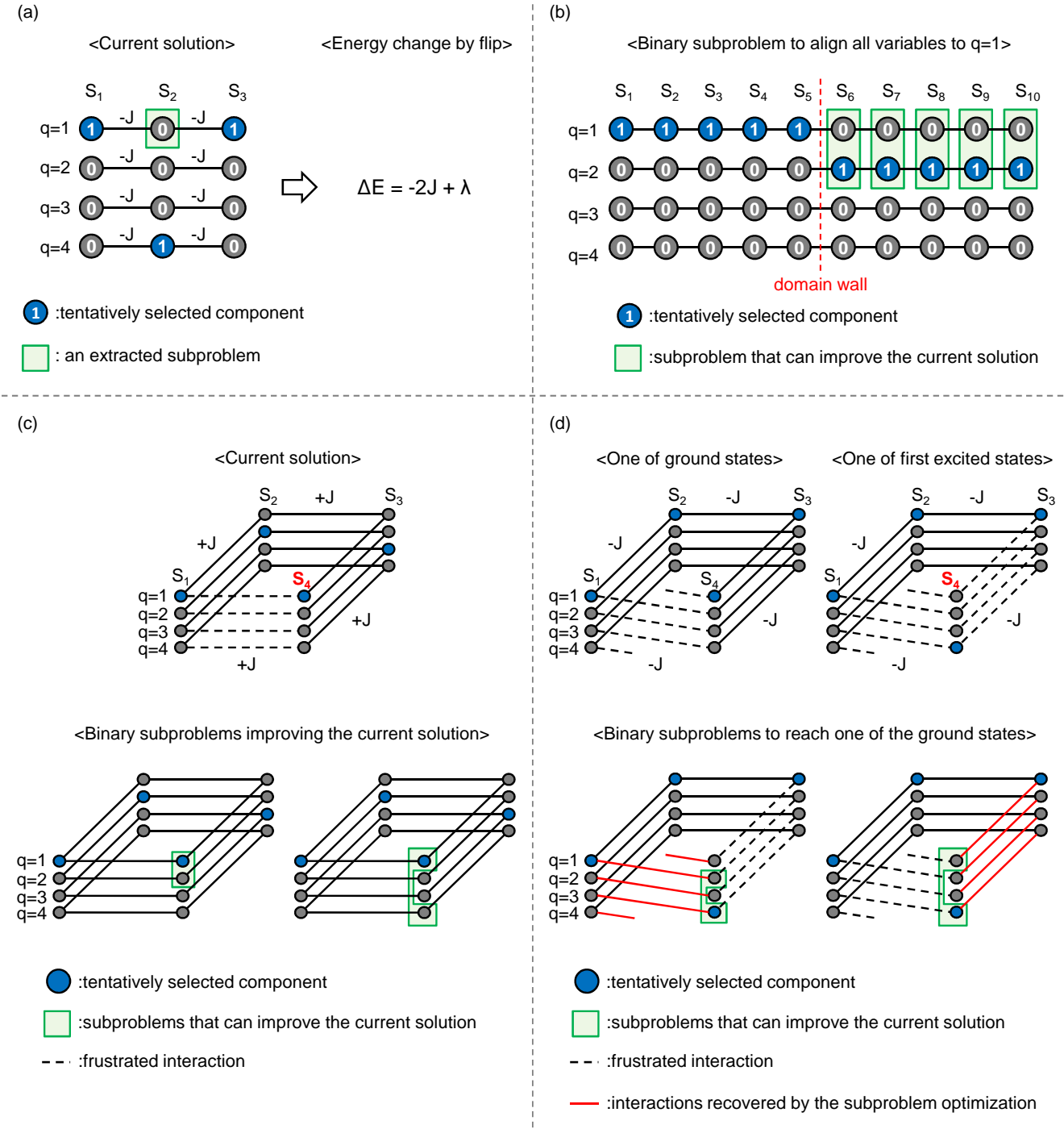}
	\caption{Discussion on results.
	Vertices and edges represent the binary variables $\{ x^{(q)}_{i} \}$ and interactions between them, respectively.
	The binary variables filled blue are the tentatively selected components, and binary variables enclosed green rectangles are the extracted subproblem.
	(a) A simple example of the random partition that can reduce the energy despite violating the one-hot constraint.
	(b) One of the first excited states, which commonly observed in the optimization of the ferromagnetic Potts model.
	In order to align all integer variables $\{ S_{i} \}_{i=1, 2, ..., 10}$ to $q=1$, the component $q=1$ must be selected for the variables $S_{6}, ..., S_{10}$.
	(c) Local interactions of the anti-ferromagnetic Potts model. The two binary subproblems can improve the current solution.
	(d) A simple example of the Potts gauge glass model.
	One interaction is frustrated in the ground state because of interactions between different components.
	Two binary subproblems can improve the first excited state. The frustrated interactions are represented by the dashed lines, and the interaction denoted by the red lines is recovered by optimizing the binary subproblem.}
	\label{fig:discussion}
\end{figure}

One of the possible answers to the first question is the existence of lower-energy unfeasible solutions in neighborhoods of the current feasible solution, and that better feasible solutions can be reached via unfeasible solutions.
A simple example of the one-dimensional ferromagnetic Potts model is shown in Fig. \ref{fig:discussion}(a).
Assume that the binary variable enclosed by the green rectangle is extracted as a one-variable subproblem, which is one of the simplest cases of the random partition.
The energy change caused by flipping the extracted binary variable is $-2J + \lambda$ because two interactions are simultaneously recovered ($-2J$) and the one-hot constraint is violated ($+ \lambda$).
If $\lambda < 2J$, flipping the binary variable decreases the energy despite violating the constraint.
Note that $\lambda > J$ is sufficient to correctly encode ground states of the one-dimensional ferromagnetic Potts model because the energy of the lowest-energy unfeasible states, where two components are commonly selected for each $S_{i}$, is $-2NJ + N\lambda$ and must be larger than that of the ground states ($-NJ$).
Consequently, if $\lambda$ is appropriately tuned ($J < \lambda < 2J$), the current solution is updated to the unfeasible solution by optimizing the subproblem; moreover, better feasible solutions will be reached via the unfeasible solution.
The number of simultaneously recovered interactions significantly contributes to the existence of such lower-energy unfeasible solutions, and it will increase as the number of frustrated interactions in ground states decreases.
Therefore, the multivalued partition is not effective in improving solution accuracy for the ferromagnetic and anti-ferromagnetic Potts models without frustrations.
Furthermore, for the Potts glass and Potts gauge glass models, the performance of the multivalued partition is better than that of the random partition because many interactions are frustrated even in ground states.

The answer to the second question is that subproblems that can eliminate domain walls are rarely extracted by the binary partition.
Figure \ref{fig:discussion}(b) shows one of first excited states which is commonly observed in the optimization of the ferromagnetic Potts model.
The ten variables in Fig. \ref{fig:discussion}(b) are divided into two domains: the five variables $S_{1}, ..., S_{5}$ are aligned to $q=1$ in one domain, whereas the other variables $S_{6}, ..., S_{10}$ are aligned to $q=2$ in the other domain.
The boundary of the domains is referrd to as domain wall.
To improve the current solution, an extracted subproblem must contain one of the ground states because the current solution is the first excited state.
For example, to align all integer variables $\{ S_{i} \}_{i=1, 2, ..., 10}$ to $q=1$, the component $q=1$ must be selected for the variables $S_{6}, ..., S_{10}$.
The probability of the component $q=1$ being selected for $S_{6}, ..., S_{10}$ is equal to $(1/3)^{5} = 1/243$ because, in addition to the tentatively selected component, the binary partition randomly selects one component for each $S_{i}$.
This probability exponentially decreases with respect to the number of variables, and the extraction of only two components is not suitable for the ferromagnetic Potts model.
Furthermore, it is conjectured that the binary partition exhibits poor performance for optimization problems that contain partial ferromagnetically ordered domains, and the concomitant use of the binary and multivalued partitions might be preferred for such problems.

The answer to the third question is that there exist several binary subproblems that can improve the current solution.
Local interactions of the anti-ferromagnetic Potts model are shown in Fig. \ref{fig:discussion}(c).
The current solution is one of the first excited states, where the interaction between $S_{1}$ and $S_{4}$ is frustrated.
Assume that the integer variable $S_{4}$ is updated to improve the current solution, then, there are two binary subproblems that can improve the current solution, which are more likely to be extracted compared to other binary subproblems.
Therefore, the disadvantage of the binary partition, which is that only two components are considered for each integer variable, is mitigated for optimizing the anti-ferromagnetic Potts model.
Hence we can exploit the advantages of the binary partition, i.e., the extracted subproblems contain a larger number of feasible solutions and the adjustment of the parameter $\lambda$ is not required.
This is also the case for the Potts glass and Potts gauge glass models, in which frustrated ground states generate several binary subproblems that improve the current solution.
Figure \ref{fig:discussion}(d) shows a simple example for the Potts gauge glass model.
One of the ground states and first excited states are shown in the top of Fig.  \ref{fig:discussion}(d), where the interaction depicted by the dashed line represents the frustrated interaction.
One interaction is frustrated in the ground state, which is caused by the interaction between different components, and two interactions are frustrated in the first excited state.
Assume that we update the integer variable $S_{4}$ to improve the current solution in the first excited state, then, there are two binary subproblems that can improve the current solution, as shown in the bottom of Fig. \ref{fig:discussion}(d):
one recovers the interaction between $S_{1}$ and $S_{4}$ and the other recovers the interaction between $S_{3}$ and $S_{4}$.
The frustrated ground states generate two binary subproblems that improve the current solution, each of which recovers different interactions.
Thus, the disadvantage of the binary partition is mitigated as long as $Q$ is not extensively large.
Note that, while the number of binary subproblems that improve the current solution increases with increasing $Q$ for the anti-ferromagnetic Potts model, it does not change for the Potts gauge glass model.

\section*{Conclusion}
We proposed two partitioning methods to efficiently solve large optimization problems under the one-hot constraint using the D-Wave quantum annealer.
The performance of the proposed methods is assessed for the ferromagnetic, anti-ferromagnetic, Potts glass, and Potts gauge glass models.
The binary partition shows the best performance among the three partitioning methods except for the ferromagnetic Potts model.
While the advantages of the binary partition are that it enables embedding a larger number of binary variables and does not require adjusting the parameter $\lambda$,
the disadvantage is that only two components are considered for each integer variable.
Although the disadvantage leads to poor performance for the ferromagnetic Potts model, it is mitigated for optimization problems that have many binary subproblems improving the current solution such as the anti-ferromagnetic Potts model and optimization problems with frustrations.
We did not identify problems for which the multivalued partition is most suitable, although the multivalued partition exhibits a better performance than the random partition for problems with frustrations.
In furure, studies should focused on constructing algorithms that can efficiently solve the ferromagnetic Potts model by the binary partition and assess the performance of the proposed methods for various optimization problems such as the graph coloring problem whose cost function is represented as the Hamiltonian of the anti-ferromagnetic Potts model.

\section*{Methods}
Details on partitioning and embedding are described in this section.

\subsection*{Multivalued Partition}
We explain how to achieve the multivalued partition using the subproblem-embedding algorithm \cite{subproblem_embed}.
The key idea of the subproblem-embedding algorithm is to exclude binary variables that are not easily embedded from the subproblem and thereby reduce the computation time.
Note that the multivalued partition requires that the binary variable, which is assigned to the tentatively selected component, must be embedded into the D-Wave quantum annealer if the integer variable is included in the subproblem.
To achieve the multivalued partition combined with the subproblem-embedding algorithm, the order of the binary variables embedded into the D-Wave quantum annealer is specified, as shown in Algorithm \ref{alg:proposed}.
We introduce two criteria to determine the order of the binary variables:
\begin{enumerate}
\item The binary variable adjacent to the already embedded binary variables.

\item The binary variable assigned to the tentatively selected component.
\end{enumerate}
After selecting an integer variable $S_{i}$ adjacent to the already embedded integer variables, the embedding order of the binary variables $\{ x^{(q)}_{i} \}_{q=1, 2, ..., Q}$ assigned to $S_{i}$ is determined according to the above criteria.
When selecting a binary variable, which is embedded into the quantum annealer first, we regard Criterion 1 as more important than Criterion 2 because embedding of the independent binary variables is a waste of the hardware resources of the quantum annealer.
For the remainder of the binary variables, we give weight to Criterion 2 to achieve the multivalued partition.
The integer variable $S_{i}$, in which only one component, is embedded is excluded from the subproblem after embedding of all binary variables $\{ x^{(q)}_{i} \}_{q=1, 2, ..., Q}$ is completed.
\begin{algorithm}
\caption{Order of binary variables embedded into the D-Wave quantum annealer}
\label{alg:proposed}
\begin{algorithmic}
\STATE{Randomly select an integer variable $S_{i}$ adjacent to the already embedded integer ones, and extract binary variables $\{ x^{(q)}_{i} \}_{q=1, 2, ..., Q}$ assigned to $S_{i}$}

\STATE{Select the binary variables that meet Criterion 1}
\IF{The binary variable satisfying Criterion 2 exists in the selected binary variables}
\STATE{Embed the binary variable into the D-Wave quantum annealer first}
\ELSE
\STATE{Randomly select one binary variable from the selected ones, and embed the binary variable first}
\ENDIF

\WHILE{Embedding of all binary variables $\{ x^{(q)}_{i} \}_{q=1, 2, ..., Q}$ is not completed}
\IF{There remains a binary variable that meets Criterion 1}
\STATE{Embed the binary variable second}
\ELSIF{There remain binary variables that meet Criterion 2}
\STATE{Embed the binary variables next}
\ELSE
\STATE{Embed other binary variables}
\ENDIF
\ENDWHILE
\end{algorithmic}
\end{algorithm}

The average number $N_{S} ( Q_{\mathrm{embed}} ) $ of embedded integer variables with $Q_{\mathrm{embed}}$ components are shown in Table \ref{tab:distribution_Qembed}, which is assessed for embedding of the Potts gauge glass model into D-Wave 2000Q\_2 with defects and is averaged over $1,000$ trials.
Note that, to distinguish the multivalued and binary partitions,  $Q_{\mathrm{embed}} > 2$ is required for most of the integer variables.
All four components are embedded for $65.8 \%$ of the integer variables in the subproblem, indicating that we can embed the multivalued subproblem which is distinct from the binary subproblem.
The average number of binary variables embedded into the D-Wave quantum annealer is $225$.

\subsection*{Binary Partition}
To solve large optimization problems by the binary partition, the cost function of the binary subproblem needs to be dirived from that of the original large problem.
The general form of the local energy between $S_{i}$ and $S_{j}$ is given by
\begin{equation}
H_{ij} = \sum_{q=1}^{Q} \sum_{q'=1}^{Q} Q^{(qq')}_{ij} x^{(q)}_{i} x^{(q')}_{j} + \sum_{q=1}^{Q} \left( Q^{(qq)}_{ii} x^{(q)}_{i} + Q^{(qq)}_{jj} x^{(q)}_{j} \right) + \lambda \left( \sum_{q=1}^{Q} x^{(q)}_{i} - 1 \right)^{2} + \lambda \left( \sum_{q=1}^{Q} x^{(q)}_{j} - 1 \right)^{2},
\end{equation}
where $Q^{(qq')}_{ij}$ represents the interaction between $x^{(q)}_{i}$ and $x^{(q')}_{j}$.
The binary partition extracts a binary subproblem by randomly selecting one component in addition to the tentatively selected component for each integer variable.
The local energy of the binary subproblem in the QUBO form is given as follows:
\begin{gather}
H_{ij}^{(\mathrm{Binary})} = R_{ij} y_{i} y_{j} + R_{ii} y_{i} + R_{jj} y_{j},  \\
R_{ij} = Q^{(\alpha_{i} \alpha_{j})}_{ij} - Q^{(\alpha_{i} \beta_{j})}_{ij} - Q^{(\beta_{i} \alpha_{j})}_{ij} + Q^{(\beta_{i} \beta_{j})}_{ij},  \\
R_{ii} = \sum_{k \neq i} \left( Q^{(\beta_{i} \alpha_{k})}_{ik} - Q^{(\alpha_{i} \alpha_{k})}_{ik} \right) - Q^{(\alpha_{i} \alpha_{i})}_{ii} + Q^{(\beta_{i} \beta_{i})}_{ii},  \\
R_{jj} = \sum_{k \neq j} \left( Q^{(\alpha_{k} \beta_{j})}_{kj} - Q^{(\alpha_{k} \alpha_{j})}_{kj} \right) - Q^{(\alpha_{j} \alpha_{j})}_{jj} + Q^{(\beta_{j} \beta_{j})}_{jj},
\end{gather}
where $y_{i} \in \{ 0, 1 \}$, $\alpha_{i}$ and $\beta_{i}$ denote the tentatively selected component and the randomly selected component for $S_{i}$, respectively, and $y_{i} = 0 (y_{i} = 1)$ indicates "stay in the tentatively selected component $\alpha_{i}$" ("transit to the other component $\beta_{i}")$.
Note that the cost function of the binary subproblem does not contain the penalty term because all solutions in the binary subproblem satisfy the one-hot constraint.

The problem graph of the binary subproblem extracted from the three-dimensional Potts model is the cubic lattice with bond dilutions.
The density of the interactions in the binary subproblem is lower than that of the multivalued subproblem because the cost function of the binary subproblem does not contain the penalty term, which generates partially fully connected interactions between $x^{(q)}_{i}$ and $x^{(q')}_{i}$.
The average number of embedded binary variables is $408$ when the binary partition is used, while it is only $225$ binary variables when the multivalued partition is used.
Furthermore, all solutions in the binary subproblem satisfy the one-hot constraint, while the multivalued subproblem does not.
Therefore, the average number $N_{\mathrm{feasible}}$ of feasible solutions involved in the embedded subproblem is considerably increased using the binary partition.
Table \ref{tab:feasible_sol} shows $\log_{10} N_{\mathrm{feasible}}$ in an embedded subproblem by using the multivalued and binary partitions combined with the complete graph embedding \cite{Comp_Embed3} and subproblem-embedding algorithm \cite{subproblem_embed}.
\begin{table}
\begin{center}
\caption{Average number of embedded integer variables with $Q_{\mathrm{embed}}$ components}
\begin{tabular}{ccc}  \hline
\begin{tabular}{c}
$Q_{\mathrm{embed}}$
\end{tabular}  &
\begin{tabular}{c}
$N_{S} ( Q_{\mathrm{embed}} )$
\end{tabular}  \\  \hline
2 & $14.5$  \\
3 & $8.0$  \\
4 & $43.2$  \\  \hline
\end{tabular}
\label{tab:distribution_Qembed}
\end{center}
\end{table}

\begin{table}
\begin{center}
\caption{Average number of feasible solutions in an embedded subproblem: $\log_{10} N_{\mathrm{feasible}}$}
\begin{tabular}{ccc}  \hline
   & Complete graph embedding & Subproblem-embedding algorithm  \\  \hline
Multivalued partition & 9.6 & 33.9  \\
Binary partition & 19.3 & 122.8  \\ \hline
\end{tabular}
\label{tab:feasible_sol}
\end{center}
\end{table}

\bibliography{sample}

\section*{Acknowledgements}
The authors are deeply grateful to Shu Tanaka, Masamichi J. Miyama and Tadashi Kadowaki for fruitful discussions.
The author M. O. is grateful for the financial support provided by JSPS KAKENHI 19H01095 and 16H04382, Next Generation High-Performance Computing Infrastructures and Applications R\&D Program by MEXT.

\section*{Author contributions statement}
S. O. conceived and developed the concept, and carried out all the experiments. M. O. proposed the plan to evaluate the validity of the concept, discussed the details of the results, and reviewed the manuscript.
S. T. directed the project in our study.

\section*{Additional information}
\textbf{Competing interests:} The authors declare no competing interests.
\end{document}